\documentclass[12pt,a4paper,oneside]{extarticle}
\usepackage{amsfonts}
\usepackage{amstext}
\usepackage{amsmath}
\usepackage{amssymb}
\usepackage[T2A]{fontenc}
\usepackage[utf8]{inputenc}
\usepackage[english]{babel}
\usepackage[numbers,square,sort]{natbib}
\bibpunct{(}{)}{,}{a}{,}{,}
\usepackage[pdftex]{graphicx}
\usepackage{lscape}
\usepackage{soul}
\usepackage{xcolor}

\usepackage{subfigure}
\usepackage{captcont}

\usepackage{geometry}
\makeatletter
\renewcommand{\@biblabel}[1]{#1.}
\makeatother

\makeatletter
 \def\capfigure{figure}
 
 \long\def\@makecaption#1#2{%
 \vskip\abovecaptionskip
 \ifx\@captype\capfigure
 \centering #1.~~#2 \par
 \else
 \raggedleft #1. \par \centering #2 \par
 \fi
 \vskip \belowcaptionskip}
 \renewcommand{\@biblabel}[1]{#1}
 \makeatother

\newtheorem{algorithm}{Algorithm}
\usepackage{tabularx}
\usepackage{booktabs}

\begin{document}

\title{Testing for explosive bubbles: a review\footnote{The author is grateful to Eiji Kurozumi, Rob Taylor, Shuping Shi, Yang Zu, Madina Karamysheva, Rustam Ibragimov, Artem Prokhorov and participants of the Center for Econometrics and Business Analysis (CEBA, St. Petersburg State University) seminar series for helpful comments and suggestions. Research was supported by a grant from the Russian Science Foundation (RSF, Project No. 20-78-10113). }}
\author{
Anton Skrobotov \\
{\small { Russian Presidential Academy of National Economy and Public Administration}}\\
{\small { Saint Petersburg State University, Center for Econometrics and Business Analysis}}}
\date{May 2022}
\maketitle


\begin{abstract}
This review discusses methods of testing for explosive bubbles in time series. A large number of recently developed testing methods under various assumptions about innovation of errors are covered. The review also considers the methods for dating explosive (bubble) regimes. Special attention is devoted to time-varying volatility in the errors. Moreover, the modelling of possible relationships between time series with explosive regimes is discussed.
\medskip

\indent \emph{Keywords}: rational bubble; testing for explosive bubble; explosive autoregression; time-varying volatility; right-tailed unit root testing.

\medskip

\indent \emph{JEL Codes}: C12, C22

\end{abstract}

\section{Introduction}

The identification of rational bubbles has been explored in a substantial body of recent research. This is explained by the causal relationship between speculative bubbles and crises in banking systems, as well as subsequent macroeconomic recessions. The most popular approach is the rational bubble model which explains explosive behaviour in prices of financial assets. In other words, following \citet{PWY2011} (hereafter PWY), consider a rational bubble by using present value theory, in which the fundamental price of the asset is the sum of present discounted values of expected future dividend. By using the no arbitrage condition
\begin{equation}\label{2}
    P_t=\frac{1}{1+R}E_t(P_{t+1}+D_{t+1}),
\end{equation}
where $P_t$ is the observed real price of an asset, $D_t$ is the observed real dividend (received from the asset for ownership between $t-1$ and $t$), and $R$ is the real interest rate used for discounting expected future cash flows. Using a log-linear approximation, as in both PWY and \citet{CS1989}, the following solution is obtained:
\begin{equation}\label{3}
    p_t=p_t^f+b_t,
\end{equation}
where
\begin{equation}\label{4}
    p_t^f=\frac{\kappa-\gamma}{1-\pi}+(1-\pi)\sum_{i=0}^{\infty}{\pi^iE_td_{t+1+i}},
\end{equation}
\begin{equation*}
    b_t=\lim_{t\rightarrow\infty}{\pi^iE_tp_{t+i}},
\end{equation*}
\begin{equation}\label{5}
    E_t(b_{t+1})=\frac{1}{\pi}b_t=(1+\exp(\overline{d-p}))b_t,
\end{equation}
with $p_t=\log(P_t)$, $d_t=\log(D_t)$, $\gamma=\log(1+R)$, $\pi=1/(1+\exp(\overline{d-p}))$, where $(\overline{d-p})$ is the average log dividend-price ratio, $\kappa=-\log(\pi)-(1-\pi)\log(1/\pi-1)$.

Therefore, the asset price in equation \eqref{3} is composed of a fundamental term $p_t^f$ (explained by expected dividends) and a rational bubble term $b_t$. For $\exp(\overline{d-p})>0$, the rational bubble $b_t$ is a sub-martingale and explosive in expectation. Under equation \eqref{5}, we have:
\begin{equation}\label{6}
    b_t=\frac{1}{\pi}b_{t-1}+\varepsilon_{b,t}=(1+g)b_{t-1}+\varepsilon_{b,t},
\end{equation}
where $E_{t-1}(\varepsilon_{b,t})=0$, $g=\frac{1}{\pi}-1=\exp(\overline{d-p})>0$  is the growth rate of the natural logarithm of the bubble, and $\varepsilon_{b,t}$ is a martingale difference.

Explosive behaviour can be statistically approximated by explosive autoregression of the form
\begin{equation}\label{1}
    y_t=\mu+\rho y_{t-1}+\varepsilon_t,
\end{equation}
with $\delta>1$.

If there is no bubble, i.e. $b_t=0$, equation \eqref{3} implies that $p_t$ is fully determined by $p_t^f$ and therefore by $d_t$. Then, \eqref{4} implies
\begin{equation}\label{7}
    d_t-p_t=-\frac{\kappa-\gamma}{1-\pi}-\sum_{i=0}^{\infty}{\pi^iE_t(\Delta d_{t+1+i})}.
\end{equation}
If $p_t$ and $d_t$ are both integrated, then \eqref{7} implies that they are cointegrated with cointegrating vector $(1,-1)$. However, with the presence of a bubble, \eqref{6} implies explosive behaviour in $b_t$, so the explosive behaviour will be in $p_t$ regardless of the behaviour of $d_t$. In this case, $\Delta p_t$ is also explosive, because if $p_t$ is the explosive process, then $p_t$ and $d_t$ can not be cointegrated. Based upon these results, \citet{DG1988} proposed to test for stationarity in $\Delta p_t$ or to test for cointegration between $p_t$ and $d_t$ for detecting the bubble. \citet[Section 2.2]{PY2011} discussed that explosiveness in price is sufficient evidence for bubbles under various assumptions.


However, \citet{DG1988} showed that impossibility of a downward rational bubble implies that the bubble will never begin again after collapsing. \citet{Evans1991} considered periodically collapsing bubbles and showed that the tests in \citet{DG1988} have low power to detect this type of a recurring bubble. The reason for this is that periodically collapsing bubbles imply a non-negligible probability of the bubble collapsing (i.e. explosive behavior occurs only temporarily in a small part of the entire sample) and behave as an $I(1)$ process and even an $I(0)$ process (i.e. the collapse is similar to a mean-reversion process).\footnote{\citet{tao2020model} investigated information criteria to select the best model among the unit root model, the local-to-unit-root model, the mildly explosive model and the regular explosive model. But their analysis was restricted to the models which use the full sample unit root model or the full sample explosive model, without any switching in some subsamples. \citet{tao2020model} also considered indirect inference estimation.} Moreover, both time series, $p_t$ and $d_t$, may be explosive and then they may be explosively cointegrated.\footnote{Explosive cointegration, or coexplosiveness, means that there exists a linear combination of explosive time series which does not contain explosive behaviour. See \citet{nielsen2010analysis} and \citet{engsted2012testing} for details.} Then, if $d_t$ is not explosive, finding of explosive behaviour in $p_t$ can be sufficient evidence of the presence of a bubble as the explosive behaviour originates only from $b_t$. 

This survey concentrates on unit root testing techniques for detecting and dating explosive bubbles. Table \ref{tab1} presents both classical and contemporary methods that were used to detect explosive bubbles in prior literature, grouped based on the area of application.

\begin{table}[!h]
\centering
\scriptsize
\caption{Applications of tests for explosive bubbles \label{tab1}}
\begin{tabular}{p{6cm}p{10cm}} \toprule
stock markets & \citet{PWY2011,phillips2015a,phillips2015b,shi2013specification,bohl2013hero,astill2017tests,astill2018real,breitung2013bubbles,fulop2017bayesian,guo2019testing,harvey2015recursive,harvey2017improving,harvey2018testing,pavlidis2017testing,kurozumiasymptotic,liu2019asymptotic,lin2020robust,monschang2020sup,whitehouse2019explosive,wang2017bubble,tao2019random,phillips2018financial,phillips2019detecting,chen2019stock}\\
prices of cryptocurrencies & \citet{cheah2015speculative,cheung2015crypto,corbet2018datestamping,corbet2019cryptocurrencies,bouri2019co,astill2020cusum,harvey2020sign,HLZ2020}\\
real estate market (housing prices) & \citet{PY2011,yiu2013detecting,anundsen2016bubbles,engsted2016explosive,pavlidis2016episodes,kivedal2013testing,das2011international,escobari2016date,shi2017speculative,caspi2016testing,banerjee2020probabilistic,harvey2020date,kurozumi2020asymptotic,pedersen2020testing,horie2016testing,chen2017inference,chen2019common,shi2020diagnosing}\\
commodities prices & \citet{gutierrez2013speculative,etienne2014bubbles,etienne2015price,zhang2016interpreting,shi2012application,fantazzini2016oil,su2017will,figuerola2015testing,caspi2018date,HLST2016,pavlidis2018using,evripidou2020cobubble}\\
exchange rates & \citet{bettendorf2013there}\\
art market & \citet{kraussl2016there}\\
long annual ratio of the US Debt/GDP series & \citet{kaufmannz2013bias}\\
credit risk in the European sovereign sector &\citet{phillips2020real} \\
extreme Yugoslavian hyperinflation & \citet{nielsen2010analysis}\\ \hline
  \bottomrule
\end{tabular}
\end{table}

The large body of empirical research justifies the relevance of the methods we will discuss. The remainder of this survey is structured as follows. Section 2 reviews various recursive right-tailed tests for explosive bubbles. Section 3 considers tests for explosive bubbles under the assumption of time-varying volatility in the innovations. Different methods for estimating the dates of exuberance and collapse as well as monitoring methods are discussed in Section 4. Section 5 discusses asymptotic results for the autoregressive parameter of explosive processes. Section 6 describes various models of the relationship between multiple time series with potentially explosive regimes. Finally, the Conclusion discusses possible further research directions.

\section{Testing for explosive behaviour}

The tests we discuss in this section are intended for testing the unit root null hypothesis in time series against the alternative of explosive behaviour in some subsample of the series. These tests are usually based on (augmented) Dickey-Fuller-type regression. In general, explosive periods are what is investigated in price time series. But in some applications, for more accuracy, one may need to decompose the price time series into fundamental and non-fundamental parts, and test for a bubble directly in the non-fundamental component. For example, \citet{shi2017speculative} investigated the bubbles in the housing market and proposed to calculate the fundamental component of rent to price ratios based on estimates of the 5-variable VAR model of the US national market. The non-fundamental component then is the difference between log price-to-rent ratio and the calculated fundamental component. \citet{shi2020diagnosing} suggested to first estimate the predictive regression of the dividend to price ratio (in first differences) on payoffs of the asset, construct fitted values of the dependent value, and cumulate them. The resulting time series is fundamental component of the dividend to price ratio series (see also \citet{ShiPhillips2020} for details). \citet{pavlidis2017testing,pavlidis2018using} proposed another approach. They used a corrected version of the real time series based on forward (futures) asset prices and market expectations of future prices to exclude the possibility of explosiveness in market fundamentals which can take place. Anyway, whether we leave or exclude a fundamental part of the time series, we need to test for the explosive behaviour of the series of interest.


\subsection{Supremum ADF test}

\citet{PWY2011} proposed recursive tests which can detect evidence of explosive behaviour in time series $\{y_t\}$, $t=1,\dots,T$.\footnote{In contrast to the methods described below, \citet{tao2019random} and \citet{banerjee2020probabilistic} considered random coefficient autoregressive processes for which it is possible to construct probabilistic forecasts of bubbles and crashes. One can use Markov-Swithing regressions in which two regimes correspond to unit root and explosive behaviour. See \citet{hall1999detecting, shi2013specification} for details. This approach, however, allows only two regimes: unit root and explosive, and adding a third regime, stationary collapsing, would make the procedure computationally expensive. See \citet{fulop2017bayesian} also for a Bayesian perspective.} The reason of using the recursive tests is that the price behaviour is dominated by the explosive (i.e., bubble) component because it is believed that the fundamental part of the price is at most $I(1)$. Therefore, we can directly test for the bubbles in prices/dividend to price ratio, not in the non-fundamental component directly.\footnote{As discussed in \citet{ShiPhillips2020}, we could apply the tests for bubbles to level variables, not logs, if we allow for a time-varying discount factor.} 

Consider the following ADF-type regression as
\begin{equation}\label{8}
    y_t=\mu+\rho y_{t-1}+\sum_{j=1}^{k}{\phi_j\Delta y_{t-j}}+\varepsilon_t.
\end{equation}
We want to test the null hypothesis of a unit root, $H_0:\rho=1$, against the right-tailed alternative, $H_1:\rho>1$, at least in some subsample. PWY proposed a recursive evolving test which consists of expanding the sample and taking the supremum over all test statistics for each subsample. In other words, we run all regressions for $t=k+1,\dots,\lfloor \tau T\rfloor$ for all $\tau\in[\tau_0,1]$ ($\lfloor\ \rfloor$ denotes the integer part of value) with some preliminary chosen $\tau_0$,\footnote{\citet{PWY2011} use $\tau_0=0.1$ in empirical application while \citet{phillips2015a} recommend to use a more flexible choice $\tau_0=0.01+1.8/\sqrt{T}$.} $T$ is the sample size of the series. 
Consider the following ADF-type test statistic:
\begin{equation}\label{9}
    ADF_\tau=\left(\frac{\sum_{j=1}^{\tau}{e_{j-1}^2}} {\hat{\sigma}_{\tau}^2}\right)^{1/2}\left(\hat{\rho}_{\tau}-1\right),
\end{equation}
where $\hat{\rho}_{\tau}$ being the OLS estimator of $\rho$ based on regression \eqref{8} over the observations $t=k+1,\dots,\lfloor\tau T\rfloor$ (first $\lfloor \tau T\rfloor$ observations), $\hat{\sigma}_{\tau}^2$ is the corresponding variance estimator of $\sigma^2_\varepsilon$, and $e_t$ are OLS residuals from \eqref{8}\footnote{Note that $ADF_1$ is the standard full-sample ADF test statistic.}. Evidently, under the null hypothesis, 
\begin{equation}\label{10}
    ADF_\tau\Rightarrow\frac{\int_0^\tau{\tilde{W(r)}dW(r)}} {\left(\int_0^\tau{\tilde{W(r)}^2}\right)^{1/2}},
\end{equation}
where $W(r)\equiv W$ is standard Brownian motion, and $\tilde{W}=W-\frac{1}{\tau}\int_0^1{W}$. The supremum-type test statistic is
\begin{equation}\label{11}
     SADF(\tau_0):=\sup_{\tau\in[\tau_0,1]}{ADF_\tau}\Rightarrow \sup_{\tau\in[\tau_0,1]}{\frac{\int_0^\tau{\tilde{W}dW}} {\left(\int_0^\tau{\tilde{W}^2}\right)^{1/2}}}.
\end{equation}
This test statistic can be used for testing for a unit root against explosive behaviour in some subsample\footnote{\citet{kaufmannz2013bias} proposed the test based on the bias corrected estimator of $\hat{\rho}_\tau$.}.  

\citet{lui2019testing} allowed for a long memory dynamic in $\varepsilon_t$. If the series exhibit long memory, then the standard SADF test diverges to infinity at rate $n^d$, where $d\in(0,0.5)$ is the memory parameter, thus, the null hypothesis of no explosive bubble is often falsely rejected. \citet{lui2019testing} suggested to replace the estimator of $\sigma^2_\varepsilon$ in \eqref{9} by a HAR (fixed-$b$) estimator. Critical values for the SADF test depend on the estimate of the memory parameter $d$, $\hat{d}$.

\subsection{More general data generating processes}

\citet{phillips2014specification} analysed and compared the limiting theory of the PWY test under different hypotheses and model specifications. The question of whether a constant and/or linear trend should be added to regression \eqref{8} was investigated. 
Different specifications under the null are also allowed. That is, PWY assumed
\begin{equation*}
    y_t=y_{t-1}+\varepsilon_t,
\end{equation*}
while \citet{DG1988} assumed
\begin{equation*}
    y_t=\tilde{\mu}+y_{t-1}+\varepsilon_t,
\end{equation*}
so that $y_t$ has a deterministic trend if $\tilde{\mu}\neq0$. \citet{phillips2014specification} considered the general specification which allows local-to-zero constant as:
\begin{equation}\label{12}
    y_t=\tilde{\mu}T^{-\eta}+y_{t-1}+\varepsilon_t,\ \eta\geq0.
\end{equation}
Here $y_t$ has a deterministic drift of the form $\tilde{\mu}t/T^{\eta}$, whose magnitude depends on sample size and localizing parameter $\eta$. If $\eta$ tends to zero or infinity, we obtain the two limiting cases considered above.

Rewriting the model \eqref{12} as
\begin{equation}\label{13}
    y_t=\frac{\tilde{\mu}t}{T^{\eta}}+\sum_{j=1}^t{\varepsilon_j}+y_0,
\end{equation}
it can be seen that the drift is small in relation to the stochastic trend when $\eta>1/2$ and equal to or stronger than the stochastic trend when $\eta\leq1/2$. Only in the last case $\eta$ can be consistently estimated (see \citet[Appendix A ]{phillips2014specification}\footnote{Consistent estimator of $\eta_T$ is $\hat{\eta}_T=-\log|\hat{\mu}_T|/\log T$ and $\tilde{\eta}_T=-\log|\tilde{\mu}_T|/\log T$, where $\hat{\mu}_T=\sum_{t=1}^T{ty_t}/ \sum_{t=1}^T{t^2}$ and $\tilde{\mu}_T=\sum_{t=1}^T{\tilde{t}y_t}/ \sum_{t=1}^T{\tilde{t}^2}$,  $\tilde{t}=t-T^{-1}\sum_{s=1}^T{s}$ are consistent estimators of $\mu_T=\tilde{\mu}T^{-\eta}$. However, there is a second order bias of $-\log|\mu|/\log T$ for $\hat{\eta}_T$ which may be sufficiently large in finite samples.}), because the drift term is dominated by stochastic trend. In other cases, the estimators of $\eta$ usually converge to 1/2 corresponding to the rate of stochastic trend (see \citet[Appendix A]{phillips2014specification} for the proof). Unfortunately, there is no finite sample comparison of the performance of the estimator $\hat{\eta}_T$.

\citet{phillips2014specification} also noted that under the alternative hypothesis, adding a constant and/or a linear trend is not realistic for actual time series (see, however, \citet{wang2017bubble}, who considered adding a linear trend).

The limiting distributions of the ADF test under the null hypothesis were obtained for $\eta>0.5$, $\eta<0.5$, and $\eta=0.5$ (these differ from \eqref{11} because of local drift). Finite sample simulations demonstrated that for $\eta>0.5$ the differences between the asymptotic and finite sample distributions are negligible regardless of different $\eta$. This is not the case for $\eta=0.5$, due to the dominating linear trend in the series, and the differences vanishes with $\eta$ approaching to zero.

In summary, \citet{phillips2014specification} recommended to always include constant term in constructing recursive ADF tests, but to compare the actual test statistic with different critical values (for $\eta>0.5$ and $\eta<0.5$) for robustness.\footnote{\citet{sollis2016fixed} analyzed  the behaviour of PWY procedure (SADF test) when there may be a break in the drift parameter. \citet{sollis2016fixed} indicated a spurious rejection of unit root null under the break. Also, he discussed the possible solutions based on pre-estimating the break date.}

\citet{PY2009} studied the following more general data generating processes which specified the new initial value after the bubble episode, so that the new unit root period begins not from the final value of the explosive period, but from a different value:
\begin{eqnarray}\label{16}
    y_t&=&y_{t-1}\mathbb I(t<T_e )+\rho_Ty_{t-1}\mathbb I(T_e\leq t\leq T_c)\\
    && +\left(\sum_{k=T_c+1}^t{\varepsilon_k}+y^*_{T_c}\right)\mathbb I(t>T_c) +\varepsilon_t\mathbb I(t\leq T_c),\notag\\
    \rho_T&=&1+\frac{c}{T^{\alpha}}, \ c>0, \ \alpha\in(0,1),\notag
\end{eqnarray}
where $T_e=\lfloor \tau_e T\rfloor$ is the origination date of the bubble, $T_c=\lfloor \tau_c T\rfloor$ is the date of its collapse, 
and thus the period $[\tau_e,\tau_c]$ is the bubble episode, 
the periods $[1,T_e)\cup (T_c,T]$ are the normal market periods, $\mathbb I(\cdot)$ denotes an indicator function. At the moment of reinitialising, $T_c$, the process is ``jumping'' to another level $y^*_{T_c}$, which can be written as $y^*_{T_c}=y_{T_e}+y^*$ with $y^*=O_p(1)$. It is assumed that $y_0=O_p(1)$.

\citet{phillips2018financial} (see also \citet{harvey2017improving} and equations \eqref{Explosive1}-\eqref{Explosive2} further in the text) considered a more reasonable mechanism that allows for transitory collapse dynamics. So, an instantaneous collapse as in \eqref{16} may be unrealistic, and some transient dynamics may be introduced after the peak~--- the so-called collapse regime. The corresponding DGP can be written as
\begin{equation}\label{17}
    y_t=
    \begin{cases} 
    \tilde{\mu} T^{-\eta}+y_{t-1}+\varepsilon_t, & t\in[1,T_e)\cup(T_r,T]\\
   (1+\delta_{1T})y_{t-1}+\varepsilon_t, &t\in[T_e,T_c]\\
    (1-\delta_{2T})\gamma_Ty_{t-1}+\varepsilon_t, &t\in(T_c,T_r]\\
    \end{cases}
    ,
\end{equation}
where $T_r=\lfloor \tau_r T\rfloor$ denotes the end of the explosive regime or the date of market recovery, so that the period $(T_c,T_r]$ is the collapse period and the periods $[1,T_e)\cup (T_r,T]$ are the normal market periods. Also, $\delta_{1T}=c_1T^{-\alpha}$, $\delta_{2T}=c_2T^{-\beta}$, $c_1,c_2>0$ and $\alpha,\beta\in[0,1)$. The formulation of AR coefficients follow moderate deviations from unity as in \citet{PM2007a}: the coefficient $\varphi_T$ deviates towards explosive behaviour, and the coefficient $\gamma_T$ deviates towards stationary behaviour. Fortunately, PWY procedure can consistently detect the bubble for this more general DGP.

\subsection{Generalized SADF test}

\citet{phillips2015a,phillips2015b} (hereafter PSY for both papers) considered the following test statistic to account for multiple explosive regimes in time series (focusing on the $\eta>0.5$ case for the drift term as more relevant in empirical applications). Their Generalized Supremum ADF (GSADF) test is
\begin{equation}\label{Cript2}
    GSADF(\tau_0)=\sup_{\tau_2\in[\tau_0,1],\tau_1\in[0,\tau_2-\tau_0]}ADF_{\tau_1}^{\tau_2},
\end{equation}
where $ADF_{\tau_1}^{\tau_2}$ is the ADF-test statistic from \eqref{9} for sample $t=\lfloor \tau_1 T\rfloor+1,\dots,\lfloor \tau_2 T\rfloor$. In this form, $\tau_{\omega}=\tau_2-\tau_1$ is a window size. That is, for every fixed $\tau_2$ the ADF test statistic is calculated over all possible $\tau_1$ from 0 to $\tau_2-\tau_0$. The GSADF test is constructed as the supremum over all possible subsample ADF test statistics with the sample size not larger than $\tau_0$. It has the following asymptotic distribution:
\begin{equation}\label{11}
     GSADF(\tau_0)\Rightarrow \sup_{\tau_2\in[\tau_0,1],\tau_1\in[0,\tau_2-\tau_0]}{\frac{\int_{\tau_1}^{\tau_2}{\tilde{W}dW}} {\left(\int_{\tau_1}^{\tau_2}{\tilde{W}^2}\right)^{1/2}}},
\end{equation}
where $W$ is standard Brownian motion, and $\tilde{W}=W-\frac{1}{\tau_2-\tau_1}\int_{\tau_1}^{\tau_2}{W}$. PSY allows minimum subsample window as $\tau_0$ (see also footnote 5) and tabulated critical values for different $\tau_0$ and sample sizes. The SADF test previously proposed by \citet{PWY2011} is a special case of GSADF, obtained by setting $\tau_1=0$ and $\tau_2=r_{\omega}\in[\tau_0,1]$.
\citet{phillips2018financial} showed that although the GSADF procedure is designed to detect bubble behaviour, it can also detect crisis periods (see also \citet{phillips2019detecting}, \citet{phillips2020real}) which are often observed in empirical applications. GSADF outperforms SADF in terms of power because GSADF is constructed in such a way that it obtains the most explosive subperiod in maximization. The GSADF plays an important role in the subsequent discussion.

\subsection{Extensions}

There are some approaches and modifications related to SADF and GSADF tests. \citet{homm2012testing} proposed to consider the supremum of the recursive Chow test through the following regression:
\begin{equation}\label{HB1}
    \Delta \tilde{y}_t=\phi_{HB}\mathbb I(t>\lfloor \tau T\rfloor)\tilde{y}_{t-1}+e_t,
\end{equation}
where $y$ is preliminary de-meaned as $\tilde{y}_t=y_t-\bar{y}$ where $\bar{y}=T^{-1}\sum_{t=1}^Ty_t$.\footnote{The lagged values of $\Delta \tilde{y}_t$ can be added to \eqref{HB1}.} The Chow-type test statistic, $C_\tau$, is defined as $t$-ratio for $\phi_{HB}$. Then the test of \citet{homm2012testing} is defined as
\begin{equation}\label{HB2}
   HB=\sup_{\tau\in[0,1-\tau_0]}C_\tau.
\end{equation}
This statistic is actually the supremum of a sequence of backward recursive statistics. \citet{harvey2015recursive} developed local-to-unit root asymptotic distribution of the HB test as well as the SADF test and found that the HB test outperformed SADF if the explosive regime belongs to the end of the sample and does not terminate. \citet{harvey2015recursive} suggested to use a so-called union of rejection testing strategy to utilize the advantages of both tests. This strategy is based on rejection at least by one of the tests, and can be written as
\[\text{Reject $H_0$ if $\{SADF>\psi_\xi q^{SADF}_\xi \text{ or } HB>\psi_\xi q^{HB}_\xi\}$},\]
with $q^{SADF}_\xi$ and $q^{HB}_\xi$ being critical values at level $\xi$, and $\psi_\xi$ being the scaling constant intended to ensure correct (asymptotic) size of the composite procedure.

\citet{korkos2019bootstrap} extended the covariate ADF (CADF) unit root testing approach of \citet{hansen1995rethinking} for testing for explosive bubbles to improve the power of the test. In this approach, the model is generated as
\begin{eqnarray}
y_t&=&\mu+u_t,\\
\Delta u_t &=& \delta u_{t-1}+\varepsilon_t,\\
\Phi(L)\varepsilon_t&=&b(L)^\prime\Delta x_t+\nu_t,
\end{eqnarray}
where $\Delta x_t$ is an $m$-vector of stationary covariates, and $\Phi(L)$  and $\beta(L)$ are some lag operators. It is assumed that $\Psi(L)\Delta x_{t+k_1+1}=z_t$ where $\Psi(L)$ is some autoregressive lag polynomial of order $l$.

The main idea is to add leads and lags of stationary covariates to regression \eqref{8} as
\begin{equation}\label{CSADF1}
    \Delta y_t=\mu+\delta y_{t-1}+\sum_{j=1}^{k}{\phi_j\Delta y_{t-j}}+\sum_{j=-k_1}^{k_2}{\beta_j\Delta x_{t-j}}+\varepsilon_t
\end{equation}
and calculate the $CADF_r$ statistic which is simply $t$-ratio for testing $\delta=0$ over the observations $t=k+1,\dots,\lfloor \tau T\rfloor$. The final $SCADF$ test statistic of \citet{korkos2019bootstrap} is defined as
\begin{equation}\label{CSADF2}
     SCADF(\tau_0):=\sup_{\tau\in[\tau_0,1]}{CADF_\tau}.
\end{equation}
The limiting distribution of the $CADF_r$ has the following form:
\begin{equation}
     SCADF(\tau_0)\Rightarrow \sup_{\tau\in[\tau_0,1]}{\frac{\int_{0}^{\tau}{\tilde{Q}(r)dP(r)}} {\left(\int_{0}^{\tau}{\tilde{Q}(s)^2}\right)^{1/2}}},    
\end{equation}
where $\tilde{Q(r)}=Q(r)-\frac{1}{\tau}\int_0^\tau{Q(r)}$, $Q(r)=b(1)\Psi(1)W_z(r)+W_\nu(r)$ and $P(s)=W_\nu(r)/\sigma_\nu$. It can be shown that the limiting distribution is based on a convex mixture of the standard normal and the Dickey-Fuller distribution with the nuisance parameter $\varrho^2$  (the value of $\varrho^2$ determines the weights and measures the relative contribution of the covariate $\Delta x_t$ to the error term $\varepsilon_t$). The estimator of $\varrho^2$ is given as $\hat{\varrho}^2=\hat{\sigma}^2_{\varepsilon\nu}/(\hat{\sigma}^2_{\varepsilon}\hat{\sigma}^2_{\nu})$, where $\sigma^2_{\varepsilon\nu}$, $\sigma^2_{\varepsilon}$, and $\sigma^2_{\nu}$ are, respectively, the covariance between $\varepsilon$ and $\nu$, the variance of $\varepsilon$, and the variance of $\nu$. All of them are estimated via the HAC approach. 

\citet{korkos2019bootstrap} proposed a bootstrap algorithm similar to \citet{chang2017bootstrapping} to obtain critical values for the $SCADF$ test and to avoid estimating $\varrho^2$ in each subsample.

\citet{whitehouse2019explosive} considered a GLS-based version of the PWY (SADF) test. Earlier, \citet{harvey2014asymptotic} investigated the OLS and GLS-based right-tailed unit root tests and found that in contrast to left-tailed tests, the GLS-based test has higher power when the magnitude of the initial condition of the series is large.\footnote{The initial condition of the series is the deviation of the first observation from the deterministic component of the process.} The GLS-based test follows from the auxiliary regression
\begin{equation}\label{SGLS1}
    \Delta \tilde{u}_t=\delta \tilde{u}_{t-1}+\varepsilon_t,
\end{equation}
where $\tilde{u}_{\tau,t}=y_t-z_t^\prime\tilde{\theta}$, $\tilde{\theta}$ is the OLS estimator from the (quasi) GLS regression $y_{\bar{c}}=(y_1,y_2-\bar{\rho}y_1,\dots,y_\tau-\bar{\rho}y_{\tau T-1})^\prime$ on $z_{\bar{c}}=(z_1,z_2-\bar{\rho}z_1,\dots,z_\tau-\bar{\rho}z_{\tau T-1})^\prime$, where $\bar{\rho}=1+\bar{c}/T$ and $z_t=1$ or $z_t=(1,t)^\prime$ is the deterministic component.\footnote{$\bar{c}=1.6$ in the constant case and $\bar{c}=2.4$ in the trend case were recommended following \citet{harvey2014asymptotic}.} Let $ADF\text{-}GLS_\tau$ be a simple $t$-ratio from the regression \eqref{SGLS1} for $t=1,\dots,\lfloor\tau T \rfloor$. Then the supremum GLS-based test proposed by \citet{whitehouse2019explosive} is of standard form:
\begin{equation}\label{SGLS2}
     SADF\text{-}GLS(\tau_0):=\sup_{\tau\in[\tau_0,1]}{ADF\text{-}GLS_\tau}.
\end{equation}
This test has higher local asymptotic power than the conventional SADF test when the explosive period is large relative to the full sample size (i.e., the proportion of the sample for which the data follows an explosive process is large). Moreover, the GLS-based test becomes better if the magnitude of initial condition increases. The initial condition does not affect the ranking of the two test types. \citet{whitehouse2019explosive} also proposed a union of rejection testing strategy based on two tests, $SADF\text{-}GLS(\tau_0)$ and $SADF(\tau_0)$ for both cases, with or without trend.

\subsection{Testing for end-of-sample bubble}

\citet{astill2017tests} considered the situation when the explosive bubble is both ongoing at the end of the sample, and of finite length. They adopted end-of-sample instability tests of \citet{andrews2003end} and \citet{andrews2006tests} for testing the null of no end-of-sample bubble against the bubble alternative. The model considered has the following form:
\begin{eqnarray}
y_t&=&\mu+u_t, \ t=1,\dots,T+m,\\
u_t&=&\begin{cases}
u_{t-1}+\varepsilon_t, \ t=1,\dots,T\\
\delta u_{t-1}+\varepsilon_t, \ t=T+1,\dots,T+m,
\end{cases}
\end{eqnarray}
where $\varepsilon_t$ is a mean zero, stationary and ergodic process. The series follows a unit root process before the moment $T$ and possibly explosive process during the following $m$ observations with $m$ being substantially smaller than $T$ and of finite length. The null hypothesis corresponds to no bubble, $\rho=1$, and the alternative hypothesis corresponds to a bubble during the end-of-sample, $\rho>1$. \citet{astill2017tests} noted that under the null, $\Delta y_t=\varepsilon_t$ during the full sample, while under the alternative, $\Delta y_t=\varepsilon_t$ up to time $T$ and $\Delta y_t=\Delta u_t=\delta(1+\delta)^{t-T-1}u_T+\sum_{j=0}^{t-T-1}{(1+\delta)^j\Delta\varepsilon_{t-j}}$, where $\delta=\rho-1$, and the first term, which is $O_p(T^1/2)$, dominates the second term, $O_p(1)$. By the first order Taylor series expansion of $(1+\delta)^{t-T-1}$ around $\delta=0$, $(1+\delta)^{t-T-1}\approx1+(t-T-1)\delta$, we have the following approximation:
\begin{equation}
    \Delta y_t=\delta(1-\delta)u_T+\delta^2u_T(t-T)+e_t,
\end{equation}
where $e_t$ contains the higher order terms in the Taylor series expansion and $O_p(1)$ term. Then, the instability test is simply the $t$-test for upward trend in regression of $\Delta y_t$ on a linear trend. Omitting the constant is correct for rolling sub-sample statistics because they are calculated before the moment $T$. The test statistic may be simply the numerator of the $t$-test for upward trend in regression of $\Delta y_t$ on a linear trend:
\begin{equation}\label{Andresw-test}
    S_m=\sum_{t=j+1}^{j+m}{(t-j)\Delta y_t}\text{ or }\sum_{t=j+1}^{j+m}{\sum_{s=t}^{j+m}\Delta y_s}.
\end{equation}
This is Andrews $S$ type statistic. The Andrews-Kim $R$ type statistic is defined as follows:
\begin{equation}
    R_m=\sum_{t=j+1}^{j+m}{\left(\sum_{s=t}^{j+m}\Delta y_s\right)^2}.
\end{equation}
The asymptotic size of $S_m$ and $R_m$ will not be affected by finite number of bubbles of finite length in the period before moment $T$. The critical values are obtained using sub-sampling techniques applied to the first $T$ observations.\footnote{\citet{astill2017tests} also recommended $m=10$ based on numerical simulations.}

To account for possible unconditional variance on innovations $\varepsilon$, \citet{astill2017tests} proposed a studentised White-type version of \eqref{Andresw-test}:
\begin{equation}\label{Andresw-test}
    S_m^{\star w}=\frac{S_m}{\sqrt{\sum_{t=j+1}^{j+m}((t-j)\Delta y_t)^2}}.
\end{equation}
\citet{astill2017tests} demonstrated that their methods dominate PSY for the case of short-lived end-of-sample bubble.

\section{Testing for explosive bubbles under time-varying volatility}

PWY and other papers discussed above assumed that the unconditional variance of the innovation process is stationary under both the null unit root and explosive alternative hypothesis. However, a general decline in the unconditional volatility of the shocks driving macroeconomic series has been a commonly observed phenomenon. Some classical unit root tests are severely oversized because their limiting distributions depend on a particular function, the so-called variance
profile, of the underlying volatility process (see \citet{Cavaliere2004a,CavaliereTaylor2007,CavaliereTaylor2008b,CavaliereTaylor2009b} and references therein). It should be noted that supremum based ADF-type test (PWY, PSY) are still robust to conditional heteroskedasticity as demonstrated by PSY.

\citet{HLST2016} addressed this issue in an explosive bubble context. Consider the following DGP for $\{y_t\}$ in time-varying parameter form
\begin{equation}\label{rho-t}
    y_t=(1+\delta_t)y_{t-1}+\varepsilon_t\qquad\mbox{or}\qquad \Delta y_t=\delta_t y_{t-1}+\varepsilon_t
\end{equation}
with obvious definition of $\delta_t$: $\delta_t$ may be $>0$ for an explosive regime, $<0$ for a stationary collapsing regime and $=0$ for a unit root regime. \citet{HLST2016} and subsequent papers considered local-to-unit root behaviour of $\delta_t$: $\delta_t=c_1/T$ for explosive period and  $\delta_t=c_2/T$ for stationary collapse period with $c_1>0$ and $c_2\leq0$. The process \eqref{rho-t} can be seen as different reparametrization of \eqref{17} except to behaviour of $\delta_t$. The non-stationary volatility is generated as $\varepsilon_t=\sigma_tz_t$, where $\{z_t\}$ is a martingale difference sequence with respect to natural filtration, and the volatility $\sigma_t$ is defined as $\sigma_{\lfloor sT\rfloor}= \omega(s)$ for $s\in[0,1]$, where $\omega(\cdot)\in {\cal D}$ is a non-stochastic and strictly positive function satisfying $0 < \underline{\omega} < \omega(s)<\bar{\omega} < \infty$. An assumption about the volatility function allows a general class of volatility processes, such as breaks in volatility, trending volatility, and regime switching volatility. \citet{HLST2016} demonstrated that, similar to classical unit root tests, asymptotic inference of the PWY (SADF) test will be affected by the presence of time-varying volatility: for the most natural cases of non-stationary volatility behaviour such as single and double breaks in volatility, and trending volatility, the SADF test is badly oversized so that it often spuriously rejects the null hypothesis of no bubble against an explosive alternative in some sub-period. To take into account this issue, \citet{HLST2016} use the following wild bootstrap. We note that although \citet{HLST2016} proposed their algorithm only for the SADF test, their methodology can be easily  implemented for the GSADF test.

\begin{algorithm}[Bootstrap Tests]

\qquad

\begin{quotation}
\begin{itemize}
\item[{\bf Step 1:}] Generate the vectors of bootstrap innovations as $e_t^*=w_t\Delta y_t$ for $t=2,\dots,T$ initialized at $e_1^*=0$, where $\{w_t\}_2^T$ be IID sequence of N(0,1) random variates.
\item[{\bf Step 2:}] Construct the bootstrap sample data via recursion $\Delta y_t^*=e_t^*$ for $t=2,\dots,T$ initialized at $y^*_0=0$.
\item[{\bf Step 3:}] Using the bootstrap sample, $\{y_t^*\}$, compute the bootstrap $GSADF$ statistic denoted as $GSADF^*$ exactly as was done for the original data for fixed lag length $k=0$.
\item[{\bf Step 4:}] Bootstrap $p$-values are then defined as:
$P_{GSADF,T}^{\ast }:=G_{GSADF,T}^{\ast }(GSADF^{\ast})$,
where $%
G_{GSADF,T}^{\ast }(\cdot )$ denote the conditional (on the original sample data)
cumulative distribution functions (cdfs) of $GSADF^{\ast}$. In practice, the cdfs required here will be unknown, but can be approximated in the usual way via numerical simulation.
\end{itemize}
\end{quotation}
\end{algorithm}

This algorithm allows a very general form of innovation variance. Although \citet{HLST2016} assumed that this variance is non-stochastic, bounded and displays a countable number of jumps, their approach still holds for the assumptions made in \citet{CavaliereTaylor2009b} (they allow stochastic limiting variance including, e.g., nonstationary autoregressive stochastic volatility, models with random volatility jumps, near-integrated GARCH processes, and explosive, nonstationary volatility). Note that in this algorithm, in Step 3, we set $k=0$ because the wild bootstrap scheme annihilates any weak dependence presented in $\Delta y_t$ in Step 1. However, \citet{pedersen2020testing} proposed a sieve based implementation of Step 2 and 3, similar to \citet{chang2003sieve}, which improve the size properties.

\citet{harvey2018testing} proposed a weighted least squares based modification of PWY test. Transform the model as
\begin{equation}\label{volat1}
    \frac{\Delta y_t}{\sigma_t}=\rho_t \frac{ y_{t-1}}{\sigma_t}+z_t, \ t=2,\dots,T.
\end{equation}
This regression is infeasible because we do not observe variance function $\sigma_t$. If $\sigma_t$ would be known, we could construct the supremum based test as PWY:
\begin{equation}
    SBZ(\tau_0)=\sup_{\tau\in[\tau_0,1]}BZ_{\tau},
\end{equation}
where $BZ_{r}$ is calculated from regression \eqref{volat1} over subsample $\{y_1,\dots,y_{\lfloor\tau T\rfloor}\}$
\begin{equation}
    BZ_{\tau}=\frac{\sum_{t=1}^{\lfloor \tau T\rfloor}{\Delta \tilde{y_t}\tilde{y}_{t-1}/\sigma_t^2}}{\left(\sum_{t=1}^{\lfloor \tau T\rfloor}{\tilde{y}_{t-1}^2/\sigma_t^2}\right)^{1/2}}.
\end{equation}
Here $\tilde{y_t}=y_t-y_1$ to guarantee the invariance to non-zero mean. The limit distribution under the null and local alternative depends on limiting volatility process $\omega(r)$. To make the test feasible, \citet{harvey2018testing} use non-parametric kernel smoothing estimator of $\sigma_t$:
\begin{equation}
    \hat{\sigma}_t^2=\frac{\sum_{i=2}^T{K_h\left(\frac{i-t}{T}\right)(\Delta y_i)^2}}{\sum_{i=2}^T{K_h\left(\frac{i-t}{T}\right)}},
\end{equation}
where $K_h(s)=K(s/h)/h$ and $K(\cdot)$ is a kernel function with a bandwidth parameter $h$.

Because the limiting distribution of the SBZ test still depends on the volatility function, \citet{harvey2018testing} utilise the wild bootstrap implementation of \citet{HLST2016} to guarantee control of size. Moreover, \citet{harvey2018testing} suggest to use a bootstrap-based union of rejections testing strategy because neither of the tests, SBZ and SADF, dominate each other across all volatility specifications (and SBZ displays non-monotonic power in some cases). This strategy has the form
\[\text{Reject $H_0$ if $\{SADF>\psi_\xi q_{\xi}^{SADF}\text{ or }SBZ>\psi_\xi q_{\xi}^{SBZ}$\}}.\]
or, equivalently,
\[\text{Reject $H_0$ if $U=\max\left(SADF,\frac{q_{\xi}^{SADF}}{q_{\xi}^{SBZ}}SBZ\right)>q_{\xi}^{U}$}.\]
where $\psi_\xi$ is a (asymptotic) scaling constant to ensure correct (asymptotic) size of the composite procedure,and $q_{\xi}^{U}=\psi_\xi\times q_{\xi}^{SADF}$ is an (asymptotic) critical value for $U$ test. 

To control size for this procedure, because the limiting distribution of the SADF and SBZ tests and therefore asymptotic critical values depend on the volatility function, the following modification of decision rule is used:
\[\text{Reject $H_0$ if $U=\max\left(SADF,\frac{q_{\xi}^{\ast,SADF}}{q_{\xi}^{\ast,SBZ}}SBZ\right)>q_{\xi}^{\ast,U}$},\]
where $q_{\ast,\xi}^{SBZ}$ and $q_{\ast,\xi}^{SADF}$ are the bootstrap based critical values for $SADF$ and $SBZ$ test statistics, and $q_{\ast,\xi}^{U}$ is the bootstrap based critical value for the $U$ test statistic.

\citet{harvey2020sign} proposed another method which controls size under time-varying volatility. This method is based on cumulated signs $C_t=\sum_{i=2}^t{sign(\Delta y_t)}$, $t=2, \dots,T$. The supremum sign-based test is defined as
\begin{equation}\label{sign1}
    sGSADF(\tau_0)=\sup_{\tau_2\in[\tau_0,1],\tau_1\in[0,\tau_2-\tau_0]}sADF_{\tau_1}^{\tau_2}.
\end{equation}
where $sADF_{\tau_1}^{\tau_2}$ is a $t$-ratio in the regression
\begin{equation}\label{sign2}
    \Delta C_t=\hat{\delta}(\tau_1,\tau_2)C_{t-1}+e_t
\end{equation}
over subsample from $\lfloor \tau_1T\rfloor$ to $\lfloor \tau_2T\rfloor$. I.e.,
\begin{equation}\label{sign2.1}
    sADF_{\tau_1}^{\tau_2}=\frac{\hat{\delta}(\tau_1,\tau_2)}{\sqrt{\hat{s}^2(\tau_1,\tau_2)/\sum_{t=\lfloor \tau_1 T\rfloor+1}^{\lfloor \tau_2 T\rfloor}C_{t-1}^2}}
\end{equation}
where $\hat{s}^2(\tau_1,\tau_2)=(\lfloor \tau_2T\rfloor-\lfloor \tau_1T\rfloor-1)^{-1}\sum_{t=\lfloor \tau_1T\rfloor+1}^{\lfloor \tau_2T\rfloor}{e_t^2}$. Because $sign(\Delta y_t)=sign(z_t)$ under the null hypothesis, the test is exact invariant to the volatility function $\sigma_t$. Moreover, the sGSADF test is exact invariant to the constant in DGP. If we allow a weak dependence of errors, we should just replace $sign(\Delta y_t)$ by $sign(\Delta y_t-\sum_{j=1}^k{\hat{\phi}_j(t)\Delta y_{t-j}})$, where $\hat{\phi}_j(t)$ are obtained from the following recursive regression
\[\Delta y_i=\hat{\alpha}(t)+\hat{\delta}(t)y_{i-1}+\sum_{j=1}^k{\hat{\phi}_j(t)\Delta y_{i-j}}+e_i\]
for $i=k+5,\dots,t$.

The special case of $sGSADF$ test is the $sSADF$ with the restriction $r_1=0$. Simulations of \citet{harvey2020sign} demonstrated that standard $GSADF$ and $SADF$ tests are more powerful (in terms of size-adjusted power) than $sGSADF$ and $sSADF$ tests respectively for large deviations from the null. Also, the $sGSADF$ is more powerful than the $sSADF$ in contrast to relationship between standard $GSADF$ and $SADF$.

\citet{harvey2020sign} also proposed a union of rejection testing strategy with wild bootstrap implementation with $GSADF$ and $sGSADF$ tests exactly in the same way as \citet{harvey2018testing}. To address the issue of asymmetric errors, \citet{harvey2020sign} replace $sign(\Delta y_t)$ by its recursive demeaned version as $sign(\Delta y_t)-(t-1)^{-1}\sum_{t=2}^t{sign(\Delta y_t)}$. The bootstrap algorithm is modified accordingly. 

\citet{hafner2020testing} modified wild bootstrap algorithm of \citet{HLST2016} to allow a skeweness of the distribution of the series. He replaced $w_t$ in Step 1 by $w_t=u_t/\sqrt{2}+(v_t^2-1)/2$ where $u_t\sim N(0,1)$ and $v_t\sim N(0,1)$, so that $E(w_t)=0$, $E(w_t^2)=1$ and $E(w_t^3)=1$. \citet{hafner2020testing} also use sieve-based recolouring in Step 2.


\citet{KST2020} proposed to use a transformation of the series according to the volatility behaviour in function $\sigma_t$, similar to \citet{CavaliereTaylor2008a} in the classical unit root testing context. Under non-stationary volatility assumption, the partial sum process of $\{\varepsilon_t\}$ is asymptotically characterized by the variance profile, termed by \citet{CavaliereTaylor2008a}, which is defined as
\[\eta(s):=\left(\int_0^1\omega(r)^2dr\right)^{-1}\int_0^s\omega(r)^2dr.\]
The so called (asymptotic) average innovation variance is defined as
\[\bar{\omega}^2:=\int_0^1\omega(r)^2dr.\]
Note that $\eta(s)=s$ under homoskedasticity. Then, we have the following weak convergence due to Theorem 1 of \citet{CavaliereTaylor2007}:
\begin{equation}
\label{fclt1}
\frac{1}{\sqrt{T}}y_{\lfloor rT\rfloor} 
=\frac{1}{\sqrt{T}}\sum_{t=1}^{\lfloor rT\rfloor}\varepsilon_t\Rightarrow \bar{\omega}W(\eta(r))=:\bar{\omega}W^\eta(r)\qquad(0\leq r \leq 1),
\end{equation}
for $y_{\lfloor\cdot \rfloor}$ defined in \eqref{rho-t}, where $\Rightarrow$ denote weak convergence in $D[0,1]$ and $W(\cdot)$ is a standard Brownian motion, while $W^\eta(\cdot)$ is called a variance transformed Brownian motion (Brownian motion under a modification of the time domain). In the case of a constant variance with $\sigma_t=\sigma$, we have $\eta(r)=r$ and thus $W^{\eta}(r)$ reduces to a standard Brownian motion. The time transformation is based on the variance profile $\eta(s)$. Roughly, we should take the sampling interval longer in the low volatility regime, whereas we take it shorter for large values of $\sigma_t$. 

More precisely, because the variance profile is a strictly monotonically increasing function, we have the unique inverse given by $g(s):=\eta^{-1}(s)$. Then, consider the time-transformed series $\tilde{y}_t=y_{t'}-y_{t'=0}$ with a non-decreasing sequence $t'=\lfloor g(t/T)T\rfloor$. We note that $\tilde{y}_0=y_0-y_0=0$ and $\tilde{y}_T=y_T-y_0$. As shown by (9) in \citet{CavaliereTaylor2007}, we have, under the null hypothesis,
\begin{equation}
\label{fclt:2}
T^{-1/2}\tilde{y}_{\lfloor r T\rfloor}\approx T^{-1/2} y_{\lfloor g(\lfloor r T\rfloor/T)T\rfloor}\approx T^{-1/2}y_{\lfloor g(r)T\rfloor} \Rightarrow \bar{\omega}W^{\eta}(g(r))=\bar{\omega}W(r)
\end{equation}
because $W^{\eta}(g(r))=W(\eta(g(r)))=W(r)$, and thus the time-transformed series behaves as if it were a unit root process with a constant variance. 

By taking into account \eqref{fclt:2}, \citet{KST2020} proposed the following test statistics based on the time-transformed ADF statistics:
\[
   STADF=\sup_{\tau_2\in[\tau_0,1]}TADF_{0}^{\tau_2}
\qquad\mbox{and}\qquad
   GSTADF(\tau_0)=\sup_{\tau_2\in[\tau_0,1],\tau_1\in[0,\tau_2-\tau_0]}TADF_{\tau_1}^{\tau_2},
\]
\begin{equation}\label{TT4}
\mbox{where}\quad
TADF_{\tau_1}^{\tau_2}=\frac{\tilde{y}_{\lfloor  \tau_2T\rfloor}^2-\tilde{y}_{\lfloor  \tau_1T\rfloor}^2-\bar{\omega}^2(\lfloor  \tau_2T\rfloor-\lfloor  \tau_1T\rfloor)}{2\bar{\omega}\sqrt{\sum_{t=\lfloor  \tau_1T\rfloor+1}^{\lfloor  \tau_2T\rfloor}\tilde{y}_{t-1}^2}}.    
\end{equation}
The limiting distributions of these test statistics are the same as in the case of homoskedasticity, therefore we do not need any bootstrap procedures to control the size. The transformation of the time series is based on volatility profile which can be estimated from data. For this, \citet{KST2020} utilise the approach of \cite{HLZ2020} and estimate non-parametrically time-varying autoregressive coefficient, collect residuals, and use them for estimating the variance profile. 

\citet{KST2020} performed Monte-Carlo simulations with comparison of $STADF$, bootstrap-based $SADF$ test of \citet{HLST2016}, sign-based test of \citet{harvey2020sign}, and SBZ test of \citet{harvey2018testing}, the latest two with union of rejection and bootstrap. \citet{KST2020} demonstrated that none of the tests dominates the others, but size of $STADF$ test is well controlled compared to the others.


\section{Detecting the dates of origination and collapse of the bubbles}

\subsection{Real-time monitoring and date estimation}

The $\sup{ADF}$ test considered in the previous section is unable to determine the date of origination  ($T_e$) and collapse ($T_c$) of the bubble. In this subsection, we consider the methods for estimating the dates which can be used in real time. PWY proposed estimators for these two dates as
\begin{equation}\label{19}
    \hat{\tau}_e=\inf_{s\geq \tau_0}{\{s:ADF_s>cv_{\beta_T}^{adf}(s)\}}, \ \hat{\tau}_c=\inf_{s\geq \tau_e}{\{s:ADF_s<cv_{\beta_T}^{adf}(s)\}},
\end{equation}
where $cv_{\beta_n}^{adf}(s)$ is a right-tailed critical values of $ADF_s$ (defined in \eqref{9}) corresponding to the significance level $\beta_T$. That is, the origination of the bubble is taken at the date for which the test statistic begins to exceed the critical value, and the date of collapse is taken at the date for which the test statistic subsequently falls below the critical value. The size $\beta_T$ should satisfy $\beta_T\rightarrow0$ as $T\rightarrow\infty$. This guarantee that $cv_{\beta_n}^{adf}(s)\rightarrow\infty$ under the null hypothesis of no bubble, so the method will not falsely detect the bubble if it actually do not present in the data.  \citet{PY2009} suggest to estimate the date of collapse at least after some period after the origination date as 
\begin{equation}\label{20}
    \hat{\tau}_c=\inf_{s\geq \tau_e+\frac{\log(T)}{T}}{\{s:ADF_s<cv_{\beta_n}^{adf}(s)\}}.
\end{equation}
This guarantee that the length of the bubble is economically significant, and the bubble episode is at least $O(\log(T))$.

PWY in their empirical study also used rolling regression for bubble dates estimation, that is, each regression was based on fixed length of a lower order than $T$. This method leads to the same estimate of the oridination date of the bubble, but leads to the earlier date of collapse.

\citet{PY2009} considered the asymptotic theory for the dates of origination and collapse of the bubble in terms of moderate deviation from a unit root process as in  \citet{PM2007a,PM2007b}. They established that under the null hypothesis of no explosive bubble, 
and under  $cv^{adf}_{\beta_T}\rightarrow\infty$, the probability of detecting the date of origination of a bubble is equal to zero as  $T\rightarrow\infty$, i.e. $Pr(\hat{\tau}_e\in[\tau_0,1])\rightarrow0$. 
Under the alternative hypothesis (mildly explosive process in model  \eqref{6}) the estimator $\hat{\tau}_e$ is consistent for  $\tau_e$ by assuming that  $1/cv^{adf}_{\beta_T}+cv^{adf}_{\beta_T}/T^{1-\alpha/2}\rightarrow0$. The idea of consistency of the date of the bubble origination is that observations from the explosive sub-period are included in the estimation of the autoregressive coefficient, these observations dominate ones from a unit root process before the explosive sub-period. The difference in the signal between the two periods then provides information and explains why the test procedure consistently estimates the dating. 

For practical implementation, \citet{PY2009} suggest to use the sequence of critical values  $cv^{adf}_{\beta_T}=\frac{2}{3}\log\log^2T$ (see \citet[Theorem 3.2]{PY2009} for details). This guarantee that the critical value converges to infinity (the Type I error is asymptotically negligible) with lower rate than $T^{1-\alpha/2}$.

For the estimator $\hat{\tau}_c$, under the null hypothesis of no explosive bubble similarly $Pr(\hat{\tau}_f\in[r_0,1])\rightarrow0$. Under the alternative hypothesis and assuming $1/cv^{adf}_{\beta_T}+cv^{adf}_{\beta_T}/T^{1-\alpha/2}\rightarrow0$, the estimator $\hat{\tau}_c$ is consistent for $\tau_c$. In this case,  $\hat{\delta}_T(\tau)\rightarrow_p1$ for $\tau>\tau_c$, but the limiting distribution of this estimator has second order downward bias  (i.e. bias toward stationary process). This bias is explained by the data under estimation which includes an explosive sub-regime as well as the period after a collapse resulting in the mean-reverting behaviour. The $ADF_{\tau}$ test diverges to $-\infty$ in this case.

Besides $ADF_{\tau}$ test, \citet{PY2009} also consider the coefficient based test, $ADF_{\tau}^{\delta}=\tau(\hat{\delta}_{\tau}-1)$. The asymptotic properties of this test are similar to the $ADF_{\tau}$ except for latter converges to $-\infty$ for $\tau>\tau_c$ with higher rate (under $\alpha>1/3$). The difference is due to the sensitivity of the standard error to the collapse of a moderate explosive process. Therefore in some cases the $ADF_{\tau}$ can better estimate the date of collapse than the $ADF_{\tau}^{\delta}$. This is confirmed by simulations in finite samples. At the other hand, the estimates of the origination of a bubble are similar for both tests.

To improve the finite sample properties of the estimators,  \citet{PY2011} proposed to select an initial condition (selected to the first observation in the previous setup) for initialisation of the recursive procedure based on the Schwartz-Bayesian information criterion (BIC). If the non-explosive regime moves to explosive regime, the most powerful procedure is based on the recursive statistics which are calculated by using only the observations from the explosive regime (because the observations from unit root process are not taken into account). In other words, let the estimate of origination date of a bubble, $\hat{T}_e=\lfloor \hat{\tau}_e T\rfloor$, is identified by recursive procedure  \eqref{19}. 
Let $n_{\min}$ is the number of observations in the sample  $\{y_{\hat{T}_e-n_{\min}+1},\dots,y_{\hat{T}_e}\}$. This sample can be constructed as some fraction (say, 10\%) from the  sample before  $\hat{T}_e$. Further, values of BIC for two competing models are compared: the unit root model and autoregressive model. For the former, the BIC is defined as 
\begin{equation*}
    BIC_{UR}=\log\left(\frac{\sum_{t=\hat{T}_e-n_{\min}-n_k}^{\hat{T}_e}{(\Delta y_t-\bar{y})^2}}{n_k +n_{\min}}\right)+\frac{\log(n_k+n_{\min})}{n_k+n_{\min}},
\end{equation*}
while for the latter, the BIC is defined as 
\begin{equation*}
    BIC_{AR}=\log\left(\frac{\sum_{t=\hat{T}_e-n_{\min}-n_k}^{\hat{T}_e}{(y_t-\hat{\mu}- \hat{\delta}y_{t-1})^2}}{n_k +n_{\min}}\right)+\frac{2\log(n_k+n_{\min})}{n_k+n_{\min}},
\end{equation*}
where $\bar{y}=(n_k+n_{\min})^{-1}\sum_{\hat{T}_e-n_{\min}-n_k}^{\hat{T}_e}{y_t}$, $\hat{\mu}$ and $\hat{\delta}$ are OLS estimators in the regression $y_t=\mu+\delta y_{t-1}+\varepsilon_t$, both BIC values are calculated for the sample  $\{y_{\hat{T}_e-n_{\min}-n_k+1},\dots,y_{\hat{T}_e}\}$. If the BIC value for the unit root model is higher and point estimate of $\delta$ is greater than one, we redefine the initial condition $\hat{T}_e-n_{\min}$ as $\hat{T}_e-n_{\min}-1$, that is add one observation to the beginning of the sample and again  compare the two BICs for larger samples. We repeat this procedure until the BIC for unit root model is lower than the BIC for autoregressive model. After this, we denote final initial condition as $\hat{T}_0$ and apply PWY procedure for the sample from $\hat{T}_0$ as in \eqref{20}. The resulting estimate of the origination date is defined as  $\hat{T}_e(\hat{T}_0)$. If the sample after repeated comparing the BICs becomes  $\{y_1,\dots,y_{\hat{T}_e}\}$, and the BIC for unit root model is larger than BIC for autoregressive model, we set the initial condition at $t=1$, so that the procedure becomes exactly the same as PWY, and the resulting estimate of the origination date $\hat{T}_e(\hat{T}_0)$ coincides with $\hat{T}_e$. In general case, it is expected that  $\hat{T}_e(\hat{T}_0)\leq\hat{T}_e$.

The alternative procedure of detecting the dates of a bubble was proposed by \citet{phillips2015a,phillips2015b} (hereafter PSY). Contrast to \eqref{19} and \eqref{20}, the estimates of the dates $T_e=\lfloor\tau_e T\rfloor$ and $T_c=\lfloor\tau_c T\rfloor$ are based on
\begin{equation}\label{21}
    \hat{\tau}_e=\inf_{\tau_2\in [\tau_0,1]}{\{\tau_2:BSADF_{\tau_2}(r_0)>scv_{\beta_T}(\tau_2)\}},
\end{equation}
\begin{equation}\label{22}
    \hat{\tau}_c=\inf_{\tau_2\in [\hat{\tau}_e+\delta\frac{\log(T)}{T},1]}{\{\tau_2:BSADF_{\tau_2}(\tau_0)<scv_{\beta_T}(\tau_2)\}},
\end{equation}
where $BSADF_{\tau_2}(\tau_0)$ is the backward sup ADF test defined as
\begin{equation*}
    BSADF_{\tau_2}(\tau_0)=\sup_{\tau_1\in[0,\tau_2-\tau_0]}ADF_{\tau_1}^{\tau_2}.
\end{equation*}
PSY extended their procedure for the case of multiple bubbles. They suggest to obtain the dates $T_e=\lfloor\tau_e T\rfloor$ and $T_c=\lfloor\tau_c T\rfloor$ for the second bubble just repeating the procedure above, but starting with collapse of the first bubble. More precisely, for the case of two bubbles the dates of originations and collapses are based on
\begin{equation}\label{23}
    \hat{\tau}_{1e}=\inf_{\tau_2\in [\tau_0,1]}{\{\tau_2:BSADF_{\tau_2}(\tau_0)>scv_{\beta_T}(\tau_2)\}},
\end{equation}
\begin{equation}\label{24}
    \hat{\tau}_{1c}=\inf_{\tau_2\in [\hat{\tau}_{1e}+\delta\frac{\log(T)}{T},1]}{\{\tau_2:BSADF_{\tau_2}(\tau_0)<scv_{\beta_T}(\tau_2)\}},
\end{equation}
\begin{equation}\label{25}
    \hat{\tau}_{2e}=\inf_{\tau_2\in [\hat{\tau}_{1f},1]}{\{\tau_2:BSADF_{\tau_2}(\tau_0)>scv_{\beta_T}(\tau_2)\}},
\end{equation}
\begin{equation}\label{26}
    \hat{\tau}_{2c}=\inf_{\tau_2\in [\hat{\tau}_{2e}+\delta\frac{\log(T)}{T},1]}{\{\tau_2:BSADF_{\tau_2}(\tau_0)<scv_{\beta_T}(\tau_2)\}}.
\end{equation}

PSY proved the consistency of the bubble dates estimators by assuming that $1/scv_{\beta_T}(r_2)+scv_{\beta_T}(r_2)/T^{1/2}\rightarrow0$. The PWY procedure consistently estimates the dates of the first bubble, but not the second bubble if the duration of the second bubble is shorter than that of the first bubble ($\tau_{1c}-\tau_{1e}>\tau_{2c}-\tau_{2e}$). If the duration of the second bubble is longer, then the PWY procedure detects the second bubble with some delay depending on the duration of the first bubble. PSY proposed a possible modification, called sequential PWY. This modification imply detect the first bubble by PWY procedure, and then start the PWY procedure again from the observation of the estimated date of the collapse (imply it as a first observation) to the end, and so on. One drawback of the sequential PWY procedure is when the minimum window length $\tau_0$
is larger than the distance between the termination dates of the two bubbles. Overall, the PSY detection procedure outperforms the sequential PWY procedure.

\citet{lui2019testing} extended the PSY procedure to allow a long memory behaviour of the errors. For this, he proposed a HAR (fixed-$b$) approach to correctly dating the explosive bubble.

For the case of data displaying time-varying volatility behaviour, \citet{phillips2020real} proposed a modification of the wild bootstrap algorithm of \citet{HLST2016}. Because the dating procedure is sequential, \citet{phillips2020real} also address the multiplicity issue. They considered the following algorithm.

\begin{algorithm}[Composite Bootstrap Procedure]

\qquad

\begin{quotation}

\begin{itemize}
\item[{\bf Step 1:}] Estimate regression 
\eqref{8} using the full sample period, imposing $\delta=0$, and obtain residuals $e_t$.
\item[{\bf Step 2:}] For the sample $\lfloor \tau_0 T\rfloor+T_b-1$ ($T_b$ is the number of observations in the window over which size is to be controlled) generate a bootstrap sample via recursion
\[\Delta y_t^*=\sum_{j=1}^k{\hat{\phi}_j\Delta y_{t-j}^*}+e_t^*\]
initialized at $y_i^*=y_i$ with $i=1,\dots,j+1$, $\hat{\phi}_j$ are OLS estimates obtained in the fitted regression from Step 1. The bootstrap innovations are generated as $e_t^*=w_te_t$, where $\{w_t\}$ be IID sequence of N(0,1) random variates.
\item[{\bf Step 3:}] Using the bootstrap sample $\{y_t^*\}$, compute the PSY test statistic sequence \\ $\{BSADF_{t}^*(\tau_0)\}_{t=\lfloor \tau_0 T\rfloor}^{\lfloor \tau_0 T\rfloor+T_b-1}$ and the maximum value of this test statistic sequence  $M_t^*=\max_{t\in[\lfloor \tau_0 T\rfloor,\lfloor \tau_0 T\rfloor+T_b-1]}BSADF_{t}^*(\tau_0)$.
\item[{\bf Step 4:}] Bootstrap critical values of the PSY procedure are obtained from the distribution of $M_t^*$.
\end{itemize}
\end{quotation}
\end{algorithm}

\citet{phillips2020real} in their empirical section used $T_b=24$ in monthly data which means that the empirical size is controlled over a two-year period. \citet{phillips2020real} also developed the R package \textit{psymonitor} to implement the proposed bootstrap procedure. \citet{phillips2020real} noted that the date-stamping procedure can also identify periods of crises.

Some papers are devoted to real-time detection of explosive bubbles in financial time series. \citet{astill2018real} extended the test proposed by \citet{astill2017tests} to a real-time monitoring scheme. The problem is that the \citet{astill2018real} test or the PSY test do not control size during sequential testing. The statistic of \citet{astill2018real} detects a bubble if any statistic in the monitoring period exceeds the largest value of the statistic calculated over a training period of data. \citet{kurozumi2020asymptotic} considered 
PWY and PSY monitoring test statistics\footnote{Actually, \citet{kurozumi2020asymptotic} considered a sequence of ADF statistics and a sequence of BSADF statistics.}, provided a new set of critical values for monitoring period and developed local asymptotic theory for the detecting statistics and CUSUM monitoring statistics of \citet{homm2012testing} (see also \citet{breitung2013bubbles}). \citet{kurozumi2020asymptotic} demonstrated that although the ADF-type monitoring scheme outperforms the CUSUM-type monitoring scheme under moderate deviation from a unit root asymptotic, this is not the case under local-to-unit root asymptotic. This is confirmed in finite sample simulations. It turns out that the CUSUM test outperforms the ADF-type tests when the bubble emerges early in the monitoring period and the bubble episode is short. \citet{astill2020cusum} extended the CUSUM test of \citet{homm2012testing} to allow for time-varying volatility. They suggested to replace standard variance estimator by its non-parametric counterpart. \citet{kurozumiasymptotic} investigated the asymptotic properties of the stopping time, which is the detecting date of a bubble. He obtained that the CUSUM-type test detects a bubble sooner than the ADF-type tests when the bubble emerges early in the monitoring period. \citet{kurozumiasymptotic} also proposed a union of rejection type procedure to monitor the bubble. This union of rejection includes both BSADF and CUSUM statistics.

\subsection{Bubble date estimation from historical data}

In this subsection, we consider the methods with more accurate break date estimation. However, these methods can not be useful for real-time monitoring. \citet{harvey2017improving} proposed an approach based on minimization of sum of squared residuals. They considered a more general DGP as \citet{phillips2018financial} (see \eqref{17}): 
\begin{eqnarray}
  y_t &=& \mu+u_t,  \label{Explosive1}\\
  u_t &=& 
\begin{cases}  
u_{t-1}+\varepsilon_t, & t=2,\dots,\lfloor \tau_{1,0}T\rfloor,\\
(1+\delta_{1})u_{t-1}+\varepsilon_t, & t=\lfloor \tau_{1,0}T\rfloor+1,\dots,\lfloor \tau_{2,0}T\rfloor,\\
(1-\delta_{2})u_{t-1}+\varepsilon_t, & t=\lfloor \tau_{2,0}T\rfloor+1,\dots,\lfloor \tau_{3,0}T\rfloor,\\
u_{t-1}+\varepsilon_t, & t=\lfloor \tau_{3,0}T\rfloor+1,\dots,T,\\
\end{cases}
\label{Explosive2}
\end{eqnarray}
This DGP implies a unit root process until the time $\lfloor\tau_{1,0}\rfloor\equiv\lfloor\tau_{e}\rfloor$, then followed by an explosive process until $\lfloor\tau_{2,0}\rfloor\equiv\lfloor\tau_{c}\rfloor$. After $\lfloor\tau_{2,0}\rfloor\equiv\lfloor\tau_{c}\rfloor$, there may be a stationary collapsing regime (which is interpreted as the return to normal market behavior) until the time $\lfloor\tau_{3,0}\rfloor\equiv\lfloor\tau_{r}\rfloor$. After the collapsing regime, the series follows a unit root process until the end of the sample. There are some special cases. If $\lfloor\tau_{2,0}\rfloor=\lfloor\tau_{3,0}\rfloor$, the explosive regime instantly changes to a unit root regime. If, moreover, $\tau_{2,0}=1$, the explosive regime is not terminated and continues to the end
of the sample period. If $\tau_{3,0}=1$, the collapsing regime is not terminated and continues to the end of the sample period as well. Actually, \citet{harvey2017improving} proposed to choose the bubble model depending on the restrictions: $\tau_{2,0}=1$ (Model 1), $\tau_{2,0}=\tau_{3,0}$ (Model 2), or $\tau_{3,0}=1$ (Model 2). Model 4 is unrestricted. For Model 4, we can estimate the corresponding regression of the form
\begin{equation}
    \Delta y_t={\mu}_1D_t(\tau_1,\tau_2)+\hat{\mu}_2D_t(\tau_2,\tau_3)+{\delta}_1D_t(\tau_1,\tau_2)y_{t-1}+{\delta}_2D_t(\tau_2,\tau_3)y_{t-1}+e_t,
\end{equation}
where $D_t(a,b)=\mathbb I(\lfloor aT\rfloor<t\leq\lfloor bT\rfloor)$. The Model 3 corresponds to the restriction $\tau_3=1$, the Model 2 corresponds to the restrictions $\mu_2=\beta_2=0$, and the Model 2 corresponds to the restrictions $\mu_2=\beta_2=0$ and $\tau_2=1$.

Let $SSR_j(\cdot)=\sum_{t=2}^T{\hat{e}_t}$ be the sum of squared residuals for the Model $j$ with the constraints $y_{\lfloor \tau_2T\rfloor}>y_{\lfloor \tau_2T\rfloor}$ and $y_{\lfloor \tau_2T\rfloor}>y_{\lfloor \tau_3T\rfloor}$.\footnote{These constraints guarantee that the explosive regime is upward and the collapsing regime is downward.} Then the dates $T_e=\lfloor \tau_eT\rfloor\equiv\lfloor \tau_1T\rfloor$, $T_c=\lfloor \tau_cT\rfloor\equiv\lfloor \tau_2T\rfloor$, and $T_r=\lfloor \tau_rT\rfloor\equiv\lfloor \tau_3T\rfloor$ can be estimated based on minimisation of the corresponding $SSR$ over all possible locations. Under the correct model and fixed magnitudes of $\delta_1$ and $\delta_2$, the dates of a bubble can be consistently estimated. If the type of the model is unknown, \citet{harvey2017improving} suggest to choose it by comparing the BIC criteria of the form
\[BIC=T\log[T^{-1}SSR(\cdot)]+k\log(T),\]
where $k=2+1$ for Model 1, $k=2+2$ for Model 2, $k=4+2$ for Model 3, and $k=4+3$ for Model 4. The minimum BIC approach always selects the appropriate model asymptotically and often outperforms the PSY procedure for dating the bubble in finite samples. It should be noted that the BIC approach must be performed only after the detecting a bubble (e.g. by GSADF test).

\citet{harvey2020sign} considered their sign-based test \eqref{sign1} for detecting the dates of the explosive bubble. The recursive implementation of the sign-based test \textit{a la} PSY can give the consistent estimator of the origination date (under mildly explosive bubble magnitude), but the date of the collapse can not be consistently estimated. Instead, \citet{harvey2020sign} suggested to maximize the sign-based statistic over all possible dates of a bubble as
\begin{equation}\label{sign3}
    \{\hat{\tau}_e,\hat{\tau}_c\}=\arg\max_{\tau_2\in[\tau_0,1], \tau_1\in[0,\tau_2-\tau_0]}sADF_{\tau_1}^{\tau_2},
\end{equation}
where $sADF_{\tau_1}^{\tau_2}$ is defined in \eqref{sign2.1}, $\hat{\tau}_e$ corresponds to $\tau_1$ and $\hat{\tau}_c$ corresponds to $\tau_2$ in maximization. It turns out that $\hat{\tau}_c$ is the consistent estimator for $\tau_c$, but $\hat{\tau}_e$ is not. To obtain the consistent estimator of the origination and collapse dates of the bubble, \citet{harvey2020sign} used the modified version of \eqref{sign2} replacing $\hat{s}^2(\tau_1,\tau_2)$ by $\tilde{s}^2(\tau_1,\tau_2))^\varepsilon$, where
\begin{equation}\label{correstion_sign}
    \tilde{s}^2(\tau_1,\tau_2)=\frac{\lfloor \tau_2T\rfloor\hat{s}^2(0,\tau_2)-\lfloor \tau_1T\rfloor\hat{s}^2(0,\tau_1)}{\lfloor \tau_2T\rfloor-\lfloor \tau_1T\rfloor-1}
\end{equation}
with $\varepsilon=0.01$. Consistency results hold under moderate deviation from the unit root in \eqref{Explosive1} and obviously require $\tau_2-\tau_1\geq \tau_0$.

\citet{monschang2020sup} investigated a robustness of the PSY date stamping procedure and sign-based date stamping procedure for a leverage effect (via the TGARCH model). They demonstrated via Monte-Carlo simulations that the PSY procedure gives more accurate estimates of the dates of a bubble than the sign-based procedure, while PSY frequently detects non-existing bubbles. Note that \citet{monschang2020sup} considered the standard sign-based procedure \eqref{sign3} without correction of the standard errors as in \eqref{correstion_sign}.


The approach of \citet{harvey2017improving} is difficult to implement for the case of  multiple bubbles because the number of potential models to be considered grows exponentially in the number of bubble regimes. \citet{pang2020estimating} proposed to consider a sample-splitting method for estimating the dates of  structural breaks: for one bubble model, initially the date of collapse is (consistently) estimated by minimizing the sum of squared residuals for the two-regime model, and the estimated break date $\hat{T}_2=\hat{T}_c$ is consistent for the date of collapse; then, treating $\hat{T}_2=\hat{T}_c$ as a known break point, the date of origination of the bubble is estimated for the sample before the estimated date of collapse. Although the origination break date estimator is not consistent, the corresponding break fraction is. However, \citet{pang2020estimating} do not allow unit root behaviour after the collapse regime. \citet{KurozumiSkrobotov2021} extended the model of \citet{pang2020estimating} by allowing a unit root regime after recovering stationary regime. It turned out that as in \citet{pang2020estimating}, the estimated break date in two-regime model is still consistent for the date of collapse. Considering the samples before and after the estimated date of collapse, the date of origination and the date of recovery can be estimated. The limiting results for the date of origination are similar to ones obtained in \citet{pang2020estimating}. The asymptotic of the date of recovery depends on the relative extent of the explosive bubble and stationary recovery regime. More precisely, if $\delta_1=1+c_1/T^a$ and $\delta_2=1-c_b/T^b$ in \eqref{Explosive2}, then the estimated date of recovery $\hat{T}_3=\hat{T}_r$ will be consistent only if $a<b$. Otherwise, $(1-\delta_2)T(\hat{\tau}_3-\tau_{3,0})$ has a non-degenerate limiting distribution. The simulation results of \citet{KurozumiSkrobotov2021} confirm the asymptotic results obtained. Also, \citet{KurozumiSkrobotov2021} investigated the performance of the sample-splitting approach under non-stationary volatility in the shocks. The demonstrated that if volatility increases, then $T_{c}$ and $T_{r}$ can be in a volatile regime. Thus, it would be difficult to distinguish between the shift in the parameter and large shocks.

\citet{harvey2020date} combined the approach of PSY and \citet{harvey2017improving} and proposed the two-step procedure. At the first step, the PSY procedure is applied to obtain preliminary dates of the bubble. At the second step, the full sample is split based on the preliminary dates of the bubbles (so that the subsample begins from the mid-point between the end of the previous explosive regime and the start of the current explosive regime, and subsample ends at the mid-point between the end of the current explosive regime and starts of the next explosive regime), and BIC based approach is applied for each of the parts of the split sample. \citet{harvey2020date} demonstrated via Monte-Carlo simulation that their two-step method is more accurate in comparison to the PSY procedure; the latter often estimates the origination dates of the bubbles later than they actually occur.

\citet{jiang2020fill} obtained the exact distribution of the maximum likelihood estimator of the structural break point in model with two regimes under continuous record asymptotic. In particular, they considered a model changing from a unit root regime to explosive regime. The distributions turn out asymmetric and trimodal regardless of the break date. \citet{jiang2020fill} also developed in-fill asymptotic distributions of break dates and demonstrated the performance of the in-fill asymptotic distribution over the long-span counterpart. Despite the difficulties of the obtained distribution, the authors suggested the method for constructing a confidence interval for the break date based on highest density region.

\section{Asymptotics for autoregressive parameter}

What should we do after detecting an explosive bubble and identifying the dates of its exuberance and collapse? We might want to estimate the growth rate of the bubble and then construct its confidence interval. The asymptotic theory needed for the explosive behaviour inference was developed in \citet{PM2007a,PM2007b}. Authors considered a mildly explosive process of the form:
\begin{equation}\label{14}
    y_t=\rho_Ty_{t-1}+\varepsilon_t, \ \rho_T=1+\frac{c}{k_T}, \ c>0,
\end{equation}
which initialises at $y_0=o_p(\sqrt{k_T})$, independent on $\{\varepsilon_t,t\geq1\}$, where  $\{k_T\}_{T\geq1}$ is the sequence tending to infinity so that $k_T=o(T)$ with $T\rightarrow\infty$. The model \eqref{14} does not include the constant to exclude the presence of deterministically explosive component in $y_t$. The error process may be weakly dependent as in \citet{PM2007b}, conditionally heteroskedastic as in \citet{arvanitis2018mildly}, or anti-persistent as in \citet{lui2020mildly}.

The sequence of autoregressive parameter $\rho_T=1+\frac{c}{k_T}>1$ has a local character, that is, $\rho_T\rightarrow1$ with $T\rightarrow\infty$. Such sequence tends to infinity with a slower rate than $T$, and represent moderate deviations from unity (the autoregressive roots belong to larger neighborhoods of one than conventional local to unity roots). One can consider a special case for $k_T=T^{\alpha}$, $\alpha\in(0,1)$. When $\alpha\rightarrow1$, we obtain a local to unit root behaviour while for $\alpha\rightarrow0$, we obtain stationary or explosive process. Under some regularity conditions, \citet{PM2007a} obtained the following results:
\begin{equation}\label{15}
    \frac{k_T(\rho_T)^T}{2c}(\hat{\rho}_T-\rho_T)\Rightarrow C, \text{ and } \frac{(\rho_T)^T}{(\rho_T)^2-1}(\hat{\rho}_T-\rho_T)\Rightarrow C,
\end{equation}
where $C$ is a standard Cauchy distribution. Note that under standard (fixed) asymptotics, the limiting Cauchy distribution is hold only under Gaussian errors, see \citet{Anderson1959,White1958}.\footnote{The standard $t$ statistic for $\delta_T$ has an asymptotically standard normal distribution, see \citet{guo2019testing}.} \citet{phillips2010smoothing} considered $\delta_{m,n}=1+\frac{cm}{T}>1$ and obtained the same result if $T\rightarrow\infty$ followed by $m\rightarrow\infty$.

The results \eqref{15} allow us to construct a confidence interval for $\rho_T$ as
\begin{equation*}
    \left(\hat{\rho}_T\pm\frac{(\hat{\rho}_T)^2-1}{(\hat{\rho}_T)^n}C_{\alpha}\right),
\end{equation*}
where $C_{\alpha}$ is the two-side $\alpha$-percentile of Cauchy distribution\footnote{The corresponding values are equal to $C_{0.10}=6.315$, $C_{0.05}=12.7$ and $C_{0.01}=63.65674$.}.

\citet{guo2019testing} generalized the \citet{PM2007a} results by allowing an deterministic drift term $\mu_T$ in \eqref{14} with $\mu_T\sqrt{k_T}\rightarrow\nu\in[0,\infty)$ (small drift) or $\mu_T\sqrt{k_T}\rightarrow\infty$ (large drift). Under i.i.d. errors $\varepsilon_t$, the $t$-statistic of $\delta_T$ under the null of moderate explosiveness is asymptotically standard normal regardless of the magnitude of the drift.\footnote{See also \citet{wang2015limit} with the theory of fixed $\rho_T=\rho>1$.} Under weakly dependent errors, the HAC based $t$-statistic also has standard normal limit. However, HAR-based $t$-statistic asymptotically has Student's $t$ distribution regardless of the magnitude of the drift. Finally, \citet{guo2019testing} proposed the constructiion of confidence interval for $\rho_T$ based on $t$ statistic inversion: if the $t$ test fails to reject the null value $\rho_T$, it should be included in the confidence interval. \citet{xiang2021testing} extended the approach of \citet{guo2019testing} by allowing a structural change in drift.

\citet{chan2012toward} (see also \citet{liu2019asymptotic}) proposed to construct a confidence interval for an autoregressive parameter with correct coverage based on empirical likelihood methods. Their approach is robust to all types of autoregressive processes: stationary, unit root, near unit root, moderate deviations from a unit root (towards stationary or explosive sides), or explosive. Similar robustness is achieved in \citet{MagdalinosPetrova} where the authors developed a testing procedure based on instrumental variable estimation and pre-testing for stationary or explosive direction. \citet{phillips2021estimation} considered estimation and inference of localizing parameters $c$ and $\alpha$ in the representation $\rho_T=1+c/T^\alpha$. Considering possibly explosive sample intervals in time series (just 4 years before the end of the sample), \citet{phillips2021estimation} argued that the method may be seen as an alternative of recursive bubble identification methods.

All methods above should be implemented for the sample interval with a stable autoregressive parameter. E.g., in the bubble model with unit root, explosive, and stationary part, we wish to do inference for the growth rate of the explosive period. For this, we first should identify the dates of exuberation and collapse of a bubble. These dates may be identified with error, and, in general, the estimated dates are not equal to the true dates. \citet{guo2019testing} and \citet{xiang2021testing} considered the sample just 100 periods before the highest
point of time series while PWY used the sample with the highest value of the Dickey-Fuller test statistic. \citet{guo2019testing} found the confidence intervals for the autoregressive parameter for seven of ten stock indices in range from 1.001 (lower bound) to 1.079. The degree of explosiveness is quite mild according the rule of \citet{PWY2011} that the explosive autoregressive root not greater than 1.05 means moderate explosiveness. \citet{xiang2021testing} considered the Bitcoin market capitalization in the fourth quarter of 2017 and found the confidence interval for autoregressive parameter to be [1.003, 1.061] which is wider if we do not pay attention to the structural change. 

It seems natural to exclude some boundary observations after the estimated date of exuberation and before the estimated date of collapse for estimation and inference of the explosive period. Also, in light of the results of \citet{pang2020estimating} and \citet{KurozumiSkrobotov2021}, the date of collapse is often estimated more accurately (for a wider set of parameters), so that there is no need to remove a lot of observations before the estimated date of collapse.

\section{Investigating the relationship between different bubbles}

A question whether bubbles in different markets move together or with some lags has also been addressed in the literature. \citet{PY2011} propose a test which detects whether the bubble migrates form one market to another. Consider two series, $x_t$ and $y_t$, and we want to test the migration from $x_t$ to $y_t$. Let $\theta_X(\tau)$ be an autoregressive coefficient characterizing $\{x_t\}_{t=1}^{\lfloor \tau T\rfloor}$. It can be recursively estimated from OLS regression as $\hat{\theta}_X(\tau)$ (i.e. $\theta_X(\tau)$ and $\hat{\theta}_X(\tau)$ are actually time series depending on $\tau$).

Let a date-stamping strategy detect the origination date of the bubble in $x_t$ as $T_{eX}=\lfloor \tau_{eX}T\rfloor$, and the autoregressive estimate  $\hat{\theta}_X(\tau)$ has a maximum at $T_{pX}=\lfloor \tau_{pX}T\rfloor$. It is assumed that $T_{pX}$ is not necessarily equal to $T_{eX}$, the date of collapse, in general.

Similarly, define $\hat{\theta}_Y(\tau)$, $T_{eY}=\lfloor \tau_{eY}T\rfloor$ and $T_{pY}=\lfloor \tau_{pY}T\rfloor$ for $\{y_t\}_{t=1}^{\lfloor \tau T\rfloor}$. It is assumed that $\tau_{pY}>\tau_{pX}$, and this inequality should be confirmed by date-stamping algorithm. 

Let $m=T_{pY}-T_{pX}$ be a number of observations in the interval $(T_{pY},T_{pX}]$. \citet{PY2011} formulate the null hypothesis that the coefficient $\theta_Y(\tau)$ for $y_t$ moves from a unit root behaviour before $T_{eY}$ (i.e. $\theta_Y(\tau)=1$) to a moderately explosive behaviour after $T_{eY}$ (i.e. $\theta_Y(\tau)=1+c_Y/T^{\alpha}$). The alternative hypothesis is that the value $\theta_Y(\tau)$ depends on the corresponding recursive coefficient $\theta_X(\tau)$ of $x_t$. Thus, the value $\theta_Y(\tau)$ is defined by the value  $\theta_X(\tau)$, i.e., when the series $x_t$ is collapsing, and $\theta_X(\tau)$ decreases, the bubble migrates from $x_t$ to $y_t$ and reveals the increase of $\theta_Y(\tau)$, which is larger than unity. 

Let the autoreressive parameter of $x_t$ behave like  $\theta_X(\tau)=1+c_X\left(\frac{\lfloor \tau T\rfloor-T_{pX}}{m}\right)/T$ for $\lfloor \tau T\rfloor>T_{pX}$, where the function $c_X(\cdot)<0$ to guarantee the mean reversion during the collapse. The dependence of $\theta_Y(\tau)$ on $\theta_X(\tau)$ can be expressed as:
\begin{equation}\label{21}
    \theta_Y(\tau)=1+\frac{c_Y+dc_X\left(\frac{\lfloor \tau T\rfloor-T_{pX}}{m}\right)}{T^{\alpha}},\ \lfloor \tau T\rfloor\geq \tau_{eY},
\end{equation}
i.e., after the origination the bubble, the behaviour of $x_t$ affects $y_t$ through the localising parameter  $c_X(\cdot)$. As the bubble collapses, $x_t$ returns to a near martingale behaviour, and the localising parameter $c_X(\cdot)$ affects the autoregressive parameter $\theta_Y$.

\citet{PY2011} consider a simplified linear version of \eqref{21}, where $c_X$ is a negative constant:
\begin{equation}\label{22}
    \theta_Y(\tau)-1=\beta_{0T}+\beta_{1T}(\theta_X(\tau)-1)\frac{\lfloor \tau T\rfloor-T_{pX}}{m}+\epsilon_t, \ \lfloor \tau T\rfloor=\lfloor \tau_{pX}T\rfloor+1,\dots,\lfloor \tau_{pY}T\rfloor,
\end{equation}
that is, the data covers the collapse period of $x_t$ and the growth of the bubble of $y_t$. The null hypothesis of no bubble migration is equivalent to $d=0$ in equation \eqref{21} or $\beta_{1T}=0$ in equation \eqref{22}. Under the alternative, $d<0$ ($\beta_{1T}<0$) because $dc_X(\cdot)>0$, which stimulates explosive behaviour of $y_t$. By using recursive coefficients $\hat{\theta}_X(\tau)$ and $\hat{\theta}_Y(\tau)$, the regression estimates of $\beta_{0T}$ and $\beta_{1T}$ can be obtained from regression \eqref{22}. 
\citet{PY2011} propose an asymptotically conservative and consistent test based on standardized test statistics $Z_{\beta}=\hat{\beta}_{1T}/L(m)$, where $1/L(m)+L(m)/n^{\varpi}\rightarrow\infty$ for all $\varpi>0$ and for some slowly varying function $L(m)$ like $a\log(m)$ with $a>0$, where $\lfloor \tau_{pY}T\rfloor-\lfloor \tau_{pX}T\rfloor=O(n)$. Standard normal critical values can be used, and the test rejects the null for large values of the test statistic. 
However, \citet{PY2011} caution against interpreting the test results as causal.



\citet{GMP2016} propose an another method to find the evidence of contagion from one time series to others. Their method implies to first compute recursive coefficients $\delta$ from regression \eqref{8} over the sample period. Let $i$ be the number of the series of interest. Authors recommended to use a fixed window width subsample in which the sequence $\{\hat{\delta}_{i,s}\}_{s=S}^T$  is recursively estimated by least square regression using a moving window of data of length $S$. In other words, $\hat{\delta}_{i,s}$ is estimated for the range from $t=s-S+1$ to $s$ for $s=S,S+1,\dots,T$. Then the contagion regression is
\begin{equation}\label{Cript5}
    \hat{\delta}_{i,s}=\theta_{1i}+\theta_{2i}\left(\frac{s}{T-S+1}\right)\hat{\delta}_{core,s-d}+\varepsilon_s, \ s=S,\dots,T,
\end{equation}
where $core$ denotes core series from which the bubble originates by assumption. The delay parameter $d$ captures the lag in market contagion from the core series on other series. \citet{GMP2016} allowed $d$ to be an integer from 0 to 12 and estimated by nonlinear least squares regression to achieve the largest $R^2$.

Regression \eqref{Cript5} is functional, so the functional coefficient $\theta_{2i}$ is time-varying. \citet{GMP2016} suggest to estimate this coefficient by a local level kernel regression (see their Technical Appendix for details). This allows to characterize the continuous change of the effect of the core series on the other series before the bubble, during the bubble and after the bubble. The larger estimate of $\theta_{2i}$, the stronger the effect. For the bubble example, the behavior of time-varying coefficient $\theta_{2i}$ may have the form of $\cap$, indicating increasing $\theta_{2i}$ during the exuberance of the bubble (contagion effect increases) and decreasing $\theta_{2i}$ during the collapse of the bubble (contagion effect decreases).

\citet{chen2019common} developed a new theory for testing  and date-stamping the common bubble factor in large-dimensional time series. More precisely, the dynamics for time series is generated as
\begin{equation}
    X_{it}=f_{0,t}\lambda_{0,i}+e_{it,}
\end{equation}
where $f_{0,t}$ follows a unit root process\footnote{See also \citet{horie2016testing} who allowed also the possible explosive behaviour in idiosyncratic component.}. This factor captures only fundamentals. Under the alternative hypothesis,
\begin{equation}
    X_{it}=f_{0,t}\lambda_{0,i}+f_{1,t}\lambda_{1,i}\mathbb I(t>T_e)+e_{it,}
\end{equation}
where the factor $f_{1,t}$ follows a mildly explosive autoregressive process, and $T_e=\lfloor \tau_eT \rfloor$ is the date of market exuberance, and the market exuberance lasts until the end of the sample period.

\citet{chen2019common} suggested to estimate the leading common factor by principal components. After that, the PSY procedure to identification the bubble is applied to the estimated first factor. The identified bubble in this  first factor is equivalent to a common bubble in large-dimensional time series $X_{it}$. The limiting distribution of the PSY test statistic is the same as original one despite of a factorisation, therefore this approach allows to consistently estimate the origination date of the common bubble.

There are other approaches operating the term co-exposiveness. \citet{magdalinos2009limit} developed the asymptotic theory for cointegrating systems with possibly mildly explosive variables.\footnote{ \citet{phillips2013inconsistent} and \citet{nielsen2008singular} also described the inconsistency results in vector autoregression with common explosive roots. } \citet{chen2017inference} extended the theory of \citet{magdalinos2009limit} to a continuous time framework (see also \citet{tao2019random}). \citet{chen2019stock} developed the asymptotic theory for spurious regression with a moderately explosive processes (see also \citet{lin2020robust}). \citet{nielsen2010analysis} proposed a cointegrated vector autoregressive model with single explosive root (co-explosive model), so that the cointegrating and co-explosive vectors eliminate random walk and explosively growing behaviour respectively via the following error correction model: 
\begin{equation}
    \Delta_1\Delta_\rho X_t=\Pi_1\Delta_\rho X_{t-1}+\Pi_\rho\Delta_1 X_{t-1}+\sum_{j=1}^{k-2}{\Phi_j\Delta_1\Delta_\rho X_{t-j}}+\mu+\varepsilon_t,
\end{equation}
where $\Delta_\rho X_t=X_t-\rho X_{t-1}$ and $\Delta_1 X_t=X_t- X_{t-1}$, and $\Pi_1$, $\Pi_\rho$ and $\Phi_j$ are $2\times 2$ parameter matrices. \citet{engsted2012testing} demonstrated how this model can be helpful for investigating the rational bubbles in asset prices.

\citet{evripidou2020cobubble} considered the so-called co-bubble behaviour via the following regression model:
\begin{equation}
    y_t=\mu_y+\beta_xx_{t-i}+\beta_zz_t+\varepsilon_{y,t},
\end{equation}
where $\varepsilon_{y,t}$ is a mean zero stationary I(0) process and $x_t$ is generated as a unit root process with bubble and collapsing regimes (see equations \eqref{Explosive1}-\eqref{Explosive2}). If $\beta_z=0$ and $\beta_x>0$, then $x_t$
and $y_t$ co-bubble, that is, their linear combination is I(0). If $i<0$, $i=0$ and $i>0$, then the bubble in $x_t$ precedes one in $y_t$ (bubble in $x_t$ migrates to $y_t$), the bubbles occur at the same time, and the $y_t$ bubble precedes one in $x_t$ (bubble in $y_t$ migrates to $x_t$), respectively. The variable $z_t$ is unobserved, the bubble process is generated in a similar way to $x_t$. So, if $\beta_z>0$ and $\beta_x=0$, then the bubble process in $x_t$ is not affected by one in $y_t$. Therefore, \citet{evripidou2020cobubble} proposed testing the null $H_0:\beta_x>0,\beta_z=0$ (co-bubble behaviour between $y_t$ and $x_t$) against $H_1:\beta_x=0,\beta_z>0$.

\citet{evripidou2020cobubble} suggested to use a KPSS-type statistic of the following form:
\begin{equation}
    S=\hat{\sigma}_y^{-2}t^{-2}\sum_{t=i\mathbb I(i>0)+1}^{T+i\mathbb I(i<0)}{\left(\sum_{s=i\mathbb I(i>0)+1}^t{\hat{e}_{y,s}} \right)^2},
\end{equation}
where $\hat{e}_{y,t}$ is the OLS residual from a regression of $y_t$ on a constant
and $x_t$, and $\hat{\sigma}_y^2=\\T^{-1}\sum_{t=i\mathbb I(i>0)+1}^{T+i\mathbb I(i<0)}{\hat{e}_{y,t}^2}$. This test is actually the test for I(0) for $\varepsilon_{y,t}$. To allow a time varying volatility in $\varepsilon_{y,t}$, \citet{evripidou2020cobubble} adopted a wild bootstrap implementation. The limiting distribution of the test statistic is free from precise properties of the regressor series $x_t$. The choice of the delay parameter $i$ should be made carefully, and different values should be experimented with in empirical applications.




\section{Conclusion}

This paper has reviewed the development of testing for explosive bubbles. Recently, many methods have been proposed to account for various features of data such as weak dependence, non-stationary conditional and unconditional volatility, and long memory. Other methods have addressed different specifications of the model generating the bubble. Overall, for empirical applications, the researcher needs to first test for an explosive bubble via tests with very general assumptions, and second (if a bubble is detected) estimate the dates of the bubble depending on model specification under similar very general assumptions. 

However, some important issues remain to
be addressed and could be investigated in further research. First, long memory dynamic can be accommodated in data generating processes coupled with heteroskedasticity (as in \citet{cavaliere2017quasi}). For now, only one paper \citet{lui2019testing} accounted for a long memory dynamic in the innovation error of the explosive bubble model. Second, penalised methods such as LASSO can be used for identification of bubble regimes. Indeed, the methods based on minimization of the sum of squared residual are computationally expensive, especially in multiple bubble episodes. LASSO methods have already been developed for the multiple breaks model (see \citet{qian2016shrinkage} \textit{inter alia}), and extensions for the multiple bubble model could be investigated.

\renewcommand{\bibname}{References}
\addcontentsline{toc}{chapter}{References}
\newpage

\bibliography{explosive,volatility,empirical}

\end{document}